\documentclass[runningheads]{llncs}

\usepackage{amssymb}
\usepackage{graphicx}

\usepackage{url}
\urldef{\mailsa}\path|{alfred.hofmann, ursula.barth, ingrid.haas, frank.holzwarth,|
\urldef{\mailsb}\path|anna.kramer, leonie.kunz, christine.reiss, nicole.sator,|
\urldef{\mailsc}\path|erika.siebert-cole, peter.strasser, lncs}@springer.com|    
\newcommand{\keywords}[1]{\par\addvspace\baselineskip
\noindent\keywordname\enspace\ignorespaces#1}

\usepackage{relsize}
\usepackage{turnstile}
\usepackage{amsmath}
\usepackage{url}

\usepackage{listings}

\usepackage{color}
\lstdefinelanguage{X10}%
  {morekeywords={abstract,break,case,catch,class,%
      const,continue,default,do,else,extends,false,final,%
      finally,for,goto,if,implements,import,instanceof,%
      interface,label,native,new,null,package,private,protected,%
      public,return,static,super,switch,synchronized,this,throw,%
      throws,transient,true,try,volatile,while,%
      async,atomic,when,foreach,ateach,finish,clocked,%
      type,here,%
      self,property,%
      proto,assert,%
      future,to,has,as,var,val,def,where,in,%
      value,or,await,current,any},%
   basicstyle=\normalfont\ttfamily,%\color{Red},%
   keywordstyle=\bf\ttfamily,%\color{OliveGreen},%
   commentstyle=\normalfont\ttfamily,%\color{Gray},%
   identifierstyle=\normalfont\ttfamily,%\color{Red},%
   stringstyle=\normalfont\ttfamily,%
   tabsize=4,%
   showstringspaces=false,%
   sensitive,%
   morecomment=[l]//,%
   morecomment=[s]{/*}{*/},%
   morestring=[b]",%
   morestring=[b]',%
   columns=fullflexible,%
   mathescape=false,%
   keepspaces=true,%
   showlines=false,%
   breaklines=true,%
   breakatwhitespace=true,%
   postbreak={},%
  }

\lstnewenvironment{xtenmath}
  {\lstset{language=X10,breaklines=false,captionpos=b,xleftmargin=2em,mathescape=true}}
  {}

\lstnewenvironment{xten}
  {\lstset{language=X10,breaklines=false,captionpos=b}} %,xleftmargin=2em
  {}

\usepackage[ruled]{algorithm} % [plain]
\usepackage[noend]{algorithmic} % [noend]
\usepackage{microtype}

\usepackage{pslatex}

\def\codesmaller{\small}

\newcommand{\code}[1]{\texttt{\textup{\codesmaller #1}}}
\newcommand{\smallcode}[1]{\texttt{\textup{\scriptsize #1}}}
\newcommand{\keyword}[1]{\code{#1}}

\usepackage{amssymb}

\usepackage{color}
\definecolor{light}{gray}{.75}

\newcommand\xX[1]{$\textsuperscript{\textit{\text{#1}}}$}

\newcommand{\typerule}[2]{
\begin{array}{c}
  #1 \\
\hline
  #2
\end{array}}
\newcommand{\typeax}[1]{
\begin{array}{c}
   \\
\hline
  #1
\end{array}}

\newcommand\mynewcommand[2]{\newcommand{#1}{#2\xspace}}

\mynewcommand{\unknown}{\code{?}}
\newcommand{\initsep}[0]{;} % tried \| and \dagger

\mynewcommand{\mycooked}{\S{}\lb\initsep\rb}

\mynewcommand{\this}{\keyword{this}}
\mynewcommand{\Object}{\code{Object}}
\mynewcommand{\const}{\keyword{const}} %C++ keyword
\mynewcommand{\mutable}{\keyword{mutable}} %C++ keyword
\mynewcommand{\romaybe}{\keyword{romaybe}} %Javari keyword

\def\withmmode#1{\relax\ifmmode#1\else{$#1$}\fi}
\def\alt{\withmmode{\;{\tt\char`\|}\;}}

\mynewcommand{\IP}{\code{I}}   % formal type parameter
\mynewcommand{\JP}{\code{J}}   % formal type parameter (for soundness proofs)

\mynewcommand{\Iparam}{Immutability parameter}
\mynewcommand{\iparam}{immutability parameter}
\mynewcommand{\iparams}{immutability parameters}
\mynewcommand{\Iparams}{Immutability parameters}
\mynewcommand{\Iarg}{Immutability argument}
\mynewcommand{\iarg}{immutability argument}
\mynewcommand{\iargs}{immutability arguments}
\mynewcommand{\Iargs}{Immutability arguments}
\mynewcommand{\ReadOnly}{\code{ReadOnly}}
\mynewcommand{\WriteOnly}{\code{WriteOnly}}
\mynewcommand{\None}{\code{None}}
\mynewcommand{\Mutable}{\code{Mutable}}
\mynewcommand{\Immut}{\code{Immut}}
\mynewcommand{\Raw}{\code{Raw}}

\mynewcommand{\This}{\code{This}}
\mynewcommand{\World}{\code{World}}

\mynewcommand{\OMutable}{\code{@OMutable}}
\mynewcommand{\OI}{\code{@OI}}

\mynewcommand{\InVariantAnnot}{\code{@InVariant}}

\newcommand{\func}[1]{\text{\textnormal{\textit{\codesmaller #1}}}}

\mynewcommand{\st}{\ensuremath{\mathrel{{\leq}}}} %{\mathop{\textrm{\tt <:}}}
\mynewcommand{\notst}{\mathrel{\st\hspace{-1.5ex}\rule[-.25em]{.4pt}{1em}~}}
\mynewcommand{\tl}{\ensuremath{\triangleleft}}
\mynewcommand{\gap}{~ ~ ~ ~ ~ ~}

\mynewcommand{\DA}{\texttt{DA}}
\mynewcommand{\ok}{\texttt{OK}}
\mynewcommand{\OK}{\texttt{OK}}
\mynewcommand{\IN}{\texttt{IN}}
\mynewcommand{\subterm}{\func{subterm}}
\mynewcommand{\TP}{\func{TP}} % function that returns type parameters in a type
\mynewcommand{\CT}{\func{CT}} % class table
\mynewcommand{\mtype}{\func{mtype}}
\mynewcommand{\mbody}{\func{mbody}}
\mynewcommand{\mmodifier}{\func{mmodifier}}
\mynewcommand{\fmodifier}{\func{fmodifier}}
\mynewcommand{\fields}{\func{fields}}
\mynewcommand{\cooked}{\func{cooked}}

\mynewcommand{\facc}{\func{acc}}
\mynewcommand{\fclocked}{\func{clocked}}

\mynewcommand{\bound}{\func{bound}_\Delta}
\mynewcommand{\substitute}{\func{substitute}}
\mynewcommand{\ftype}{\func{ftype}}
\mynewcommand{\mguard}{\func{mguard}}
\mynewcommand{\isTransitive}{\func{isTransitive}}

\mynewcommand{\hnull}{\code{null}}
\mynewcommand{\htrue}{\code{true}}
\mynewcommand{\hfalse}{\code{false}}

\mynewcommand{\hUnused}{\code{\_}}
\mynewcommand{\hA}{\code{A}} % inVariant definition
\mynewcommand{\hB}{\code{B}} % inVariant definition
\mynewcommand{\hC}{\code{C}} % class
\mynewcommand{\hD}{\code{D}} % class
\mynewcommand{\hF}{\code{F}} % field declaration
\mynewcommand{\hG}{\code{G}}
\mynewcommand{\hI}{\code{I}} % iparam
\mynewcommand{\hJ}{\code{J}}
\mynewcommand{\hL}{\code{L}} % class decl
\mynewcommand{\hM}{\code{M}} % Method decl
\mynewcommand{\hN}{\code{N}} % Non-variable type
\mynewcommand{\hO}{\code{O}}
\mynewcommand{\hP}{\code{P}}
\mynewcommand{\hR}{\code{R}}
\mynewcommand{\hS}{\code{S}}
\mynewcommand{\hT}{\code{T}} % types (vars or non vars)
\mynewcommand{\hU}{\code{U}} % types (vars or non vars)
\mynewcommand{\hV}{\code{V}} % closed types
\mynewcommand{\hX}{\code{X}} % vars
\mynewcommand{\hY}{\code{Y}} % vars
\mynewcommand{\hZ}{\code{Z}} % inVariant definition

\mynewcommand{\ha}{\code{a}}
\mynewcommand{\hc}{\code{c}} % cooker
\mynewcommand{\hd}{\code{d}}
\mynewcommand{\hm}{\code{m}} % method
\mynewcommand{\he}{\code{e}} % expression
\mynewcommand{\hf}{\code{f}} % field
\mynewcommand{\hg}{\code{g}}
\mynewcommand{\hl}{\code{l}} % location in the store
\mynewcommand{\ho}{\code{o}}
\mynewcommand{\hp}{\code{p}} % cooker
\mynewcommand{\hq}{\code{q}}
\mynewcommand{\hr}{\code{r}}
\mynewcommand{\hv}{\code{v}} % value
\mynewcommand{\hw}{\code{w}} % value
\mynewcommand{\hx}{\code{x}} % method parameter
\mynewcommand{\hy}{\code{y}} % field
\mynewcommand{\hz}{\code{z}} % method parameter

\mynewcommand{\lroot}{\code{l}_\top} % root
\mynewcommand{\lthis}{\code{l}_\smallcode{this}} % this

\mynewcommand{\hasync}{\code{async}}
\mynewcommand{\hswitch}{\code{switch}}
\mynewcommand{\hAcc}{\code{Acc}}
\mynewcommand{\hfinish}{\code{finish}}
\mynewcommand{\hreceiver}{\code{receiver}} % receiver for new
\mynewcommand{\hSW}{\code{SW}}
\mynewcommand{\hAW}{\code{AW}}
\mynewcommand{\hObject}{\code{Object}}

\mynewcommand{\hdef}{\code{def}}
\mynewcommand{\hfor}{\code{for}}
\mynewcommand{\hvar}{\code{var}}
\mynewcommand{\hin}{\code{in}}
\mynewcommand{\hPoint}{\code{Point}}
\mynewcommand{\hand}{\code{~and~}}
\mynewcommand{\hor}{\code{~or~}}
\mynewcommand{\hthis}{\code{this}} % this
\mynewcommand{\hclass}{\code{class}}
\mynewcommand{\hreturn}{\code{return}}
\mynewcommand{\hhnew}{\code{new}}
\mynewcommand{\hsub}{\code{/}} % substitute (reduction rules)
\mynewcommand{\nonescaping}{\code{nonescaping}}
\mynewcommand{\hescaping}{\code{escaping}}
\mynewcommand{\hextends}{\code{extends}}

\newcommand{\PREV}[1]{\withmmode{\mathtt{prev}~{#1}}}
\newcommand{\ift}[2]{\withmmode{\mathtt{if}~{#1}~\mathtt{then}~{#2}}}
\newcommand{\ife}[2]{\withmmode{\mathtt{if}~{#1}~\mathtt{else}~{#2}}}
\newcommand{\some}[2]{\withmmode{\mathtt{some}~{#1}\,\mathtt{in}\,~{#2}}}
\newcommand{\all}[2]{\withmmode{\mathtt{all}~{#1}\,\mathtt{in}\,{#2}}}
\newcommand{\hence}[1]{\withmmode{\mathtt{hence}~{#1}}}
\newcommand{\hitherto}[1]{\withmmode{\mathtt{hitherto}~{#1}}}
\newcommand{\AND}[2]{\withmmode{{#1}~\mathtt{and}~{#2}}}
\newcommand{\OR}[2]{\withmmode{{#1}~\mathtt{or}~{#2}}}
\newcommand{\MU}[2]{\withmmode{\mathtt{mu}~{#1}\ \mathtt{in}\ {#2}}}

\newcommand{\TIME}[2]{\withmmode{\mathtt{time}^{#1}\,{#2}}}

\newcommand{\proves}[3]{\withmmode{\dststile{#3}{#1,#2}}}
\newcommand{\evolves}[3]{\withmmode{\longrightarrow^{#1,#2}_{#3}}}
\newcommand{\starevolves}[3]{\withmmode{\stackrel{\star}{\longrightarrow}^{#1,#2}_{#3}}}
\newcommand{\steps}[2]{\withmmode{\leadsto^{#1,#2}}}
\mynewcommand\xth{\xX{th}}
\mynewcommand\xrd{\xX{rd}}
\mynewcommand\xnd{\xX{nd}}
\mynewcommand\xst{\xX{st}}
\mynewcommand\ith{$i$\xth}
\mynewcommand\jth{$j$\xth}

\mynewcommand{\myindent}{~~}

\makeatletter
\def\topfigrule{\kern3\p@ \hrule \kern -3.4\p@} % the \hrule is .4pt high
\def\botfigrule{\kern-3\p@ \hrule \kern 2.6\p@} % the \hrule is .4pt high
\def\dblfigrule{\kern3\p@ \hrule \kern -3.4\p@} % the \hrule is .4pt high
\makeatother


\usepackage{bold-extra}

\newcommand{\ttlcb}{\texttt{\char "7B}}
\newcommand{\ttrcb}{\texttt{\char "7D}}
\newcommand{\lb}{\ttlcb}
\newcommand{\rb}{\ttrcb}
\newcommand{\cc}{{\sf CC}}

\def\from#1\infer#2{{{\textstyle #1}\over{\textstyle #2}}}
\def\rname#1\from#2\infer#3{{{\textstyle #2}\over{\textstyle #3}}{\ \textstyle(#1)}}

\begin{document}

\mainmatter  % start of an individual contribution

\title{TCC, with History}

\titlerunning{TCC, with History}

\author{Vijay Saraswat \inst{1}
\and Vineet Gupta \inst{2}\and Radha Jagadeesan\inst{3}\thanks{Radha Jagadeesan was supported by NSF 0916741.}}
\authorrunning{TCC, with History}

\institute{IBM TJ Watson Research Center \and Google, Inc. \and DePaul University}


\toctitle{Lecture Notes in Computer Science}
\tocauthor{Saraswat, Gupta and Jagadesan}
\maketitle

\begin{abstract}

Modern computer systems are awash in a sea of asynchronous
events. There is an increasing need for a declarative language that
can permit business users to specify complex event-processing
rules. Such rules should be able to correlate different event streams,
detect absence of events (negative information), permit aggregations
over sliding windows, specify dependent sliding windows etc. For
instance it should be possible to precisely state a rule such as
``Every seventh trading session that DowJones has risen consecutively,
and IBM's stock is off $3\%$over its average in this period, evaluate
IBM position'', ``Declare the sensor as faulty if no reading has been
received for $500$ ms'', etc. Further, the language should be
implementable efficiently in an event-driven fashion.    

We propose the Timed (Default) Concurrent Constraint, TCC,
programming framework as a foundation for such complex event
processing. The framework (developed in the mid 90s) interprets
computation as deduction in a fragment of linear temporal logic. It
permits the programmer to write rules that can react instantaneously
to incoming events and determine the ``resumption" that will respond to
subsequent events. The framework is very powerful in that it permits
instantaneous pre-emption, and allows user-definable temporal
operators (``multi-form time"). 

However, the TCC framework ``forgets" information from one instant to
the next. We make two extensions. First, we extend the TCC model to
carry the store from previous time instants as ``past" information in
the current time instant. This permits rules to to be written with
rich queries over the past. Second, we show that many of the powerful
properties of the agent language can be folded into the query language
by permitting agents and queries to be defined mutually recursively,
building on the testing interpretation of intuitionistic logic
described in RCC~\cite{radha-fsttcs05}. We show that this permits
queries to move ``back and forth" in the past, e.g.{} ``Order a review if
the last time that IBM stock price dropped by $10\%$ in a day, there
was more than $20\%$ increase in trading volume for Oracle the
following day." 

We provide a formal semantics for TCC + Histories and establish
some basic properties.

\end{abstract}

\section{Introduction}\label{sec:intro}
\subsection{Timed Concurrent Constraint Programming}
From about 1985 to about 1995, the programming languages/embedded
systems community worked out a very robust programming model for
time-based systems, under the framework of ``synchronous languages'',
such as Esterel, Signal and Lustre (\cite{esterel,statecharts,signal,lustre}). In particular, the authors
developed the {\em Timed (Default) Concurrent Constraint Programming
  Framework}, \cite{saraswat-tdcc}, based on the simple idea of extending ``across time'' the
ideas of concurrent constraint programming, using the Synchrony
Hypothesis of Berry \cite{esterel}.\footnote{In the rest of this paper we will use the acronym TCC to stand for Timed Default Concurrent Constraint Programming.}
One thinks of a reactive
system as lying inert, waiting for a stimulus from the outside world.
On each stimulation, the system computes an instantaneous response,
and prepares itself for further interaction (by computing a
resumption). The system is {\em amnesiac} in that its past state is
flushed, only the resumption is kept. Thus the system has an internal
notion of time that corresponds to its periodic interaction with the
outside world.

This notion of time can be made explicit through certain temporal
combinators within the language used to program these agents.  TCC is
built on just six orthogonal basic combinators:\footnote{ We 
 introduce recursion explicitly through \code{mu}; in fact recursion
 is definable in TCC.}

\vspace{5mm}
{\footnotesize
\begin{tabular}{l}
(Agents) A,B  {::=}  c   \alt \ift{G}{A}    \alt \ife{c}{A}    \alt
\AND{A}{A} \alt \some{V}{A} \alt \hence{A}
\\ \quad \alt  Z \alt \MU{Z}{A}\\ 
(Goals) G  {::=}  c   \alt \AND{G}{G}
\end{tabular}}
\vspace{5mm}

\noindent Above, c ranges over constraints; X,V over
first-order variables used in constraints; Z over Agent variables;
A,B ranges over Agent formulas, and G over Goal formulas. 

The TCC framework is parametric on an underlying notion of {\em
  constraint system} $C$ \cite{saraswat-tdcc}: essentially such a
system specifies pieces of partial information, called {\em tokens} or
{\em constraints}, and an {\em entailment} relation which specifies
which tokens follow from which other sets of tokens. The (tell) \code{c} agent 
adds the constraint \code{c} to a shared store of constraints.
The (positive ask) agent \code{if c then A} reduces to \code{A} if 
the store is strong enough to entail \code{c}. The (negative ask)
agent \code{if c else A} reduces to 
\code{A} only if the final store (at this time instant) will not be
strong enough to entail \code{c} (this circularity -- the final store
is defined in terms of the final store -- is characteristic
of defaults \cite{Reiter1980-REIALF}).
The (parallel composition) agent \code{A and B} behaves as both
\code {A} and \code{B}. The agent \code{some X in A}
introduces a new local variable \code{X} in \code{A}. The agent
\code{hence A} is the only agent with temporal behavior -- it reduces
to \code{A} at every time instant after the current instant.   The
agent \code{mu Z A} (taken from the modal mu calculus) behaves like
\code{A} with occurrences of \code{Z} replaced by \code{mu
  Z A}.%, i.e.{} the agent \code{A [ mu Z A / Z]}.

This language is powerful enough to be the basis for a rich
algebra of temporal control constructs. For instance, one can define:  
\begin{enumerate}
\item 
\code{always A} (run \code{A} at every time step); 
\item 
\code{do A
  watching c} (run \code{A} until such time instant as the condition
\code{c} is true, at which point abort the remainder of \code{A});
\item 
\code{next A} (run \code{A} only at the next time step);  
\item 
\code{time A on c} (run \code{A} but on a clock derived from the basic
clock by only passing through those ticks at which the condition
\code{c} is true). 
\end{enumerate}

The last combinator in particular is very powerful -- it realizes the
idea of ``multi-form'' time, the notion that the basic clock on which
an agent is defined may itself be defined by another agent \cite{saraswat-tdcc}.  

TCC (and its continuous time extension, HCC, \cite{vineet-ccc}) have
been used in modeling complex electro-mechanical systems (photo
copiers \cite{Gupta95modelinga}, robots \cite{aercam}) and biological systems \cite{hcc-biology}. They have a very
well-developed  theory -- semantic foundations, reasoning framework,
implementation techniques, compilation into finite state automata,
abstract interpretation, etc. (see Related Works section below).

Unlike the other systems mentioned above (Esterel, and other reactive
languages), TCC, and its parent framework, Concurrent Constraint Programming (CCP) are  declarative
and rule-based.  Computation can be interpreted as deduction
corresponding to certain ``agent'' formulas in linear time temporal
logic, defined over a certain notion of {\em defaults}
\cite{saraswat-tdcc}.  Defaults play
a crucial role in permitting agents to detect the {\em absence}  of
information. This is critical for faithfully modeling such
computational phenomena as time-outs and strong pre-emption. 
 This logical reading extends the understanding of CCP
\cite{saraswat-thesis} as computation in intuitionistic logic
\cite{saraswat92higherorder}.

\subsection{Event processing}
Over the last decade a new and interesting application area has
emerged,  {\em event processing}, \cite{luckham-events}. The
basic computational problem in event processing is to implement a
powerful ``sense, analyze, respond'' system. The system should be
capable of receiving multiple (usually discrete) time-varying signals,
correlating them in potentially complex ways involving detecting the
absence of events, maintaining sliding windows, computing statistics
over sliding windows (averages, max values, etc), and comparing these
values. If the desired temporal pattern is detected, then appropriate
programmer specified action (e.g.{} issuing an alert) needs to be
taken. 

For an event processing language to be useful, it should be capable of
expressing complex patterns of temporal interactions. For example,
it should be possible to support rules of the form:

\begin{enumerate}

\item   Every tenth time the price drops within an hour emit volatility
    warning.

\item   Every seventh trading session that DowJones rises
  consecutively, and IBM stock has fallen over this interval, evaluate
  IBM position.

\item   Declare the sensor is faulty if no reading has been received in the
   past $500s$.

\item   Declare the room is too cold if the average temperature over the
   last $100s$ is below a threshold.

\item    Ignore an over global limit notification on an account if an over
   global limit notification was sent on this account in the past two
   days.

\item   If the merchant has been tenured less than $90$ days, and the sum of
   the transactions in the last $7$ days is much higher than the
   seven day average for the last $90$ days, then investigate a $7$ day hit
   and run possibility.
\end{enumerate}

We also desire a language in which programs can be understood
declaratively as ``rules''.  Ultimately the language needs to mesh
well with an OPS-like rule language, such as ILOG JRules and ILOG
Business Rules. We desire that the programmer should be able to reason
rigorously (if informally) about such rules.  We require that the
rules should be compilable efficiently.  For example queries involving
sliding window averages should be implemented in an incremental
forward-driven fashion (with a rolling average being maintained). 

\subsection{TCC for event processing}
Given the many valuable properties of TCC, it is interesting to
consider it as a basis for complex event processing. Incoming events
can be represented as atoms to be added to the constraint store. 
%(For
%concreteness we may consider the underlying constraint system to be
%Gentzen \cite{Saraswat94programmingin}; see Appendix~\ref{sec:Gentzen}.)
As events arrive, they are buffered while the system is active (executing
events it received at the previous tick). Once the system quiesces,
and the buffer is not empty, the system is advanced to the next time
unit, and all buffered events added.

A fundamental limitation of TCC for complex event processing,
however, is that TCC computations do not maintain history.
All rules must be written in a ``forward looking''
fashion, responding to the current events received, and whatever state
has been explicitly stored from past interactions. For instance, to
express the rule ``Trigger an alert whenver it is the case that the
stock price of company A falls over $10\%$, while that of company B
has risen over the past 7 days'', the programmer must write code that
maintains in the current store the value of the proposition ``the
stock of company B has risen over the past 7 days''. Now on receipt of
a notification that the stock price of A has fallen, a check can be
made for the value of the proposition and an alert emitted if
necessary. 

But this way of writing rules is awkward. In essence, the programmer
is being made to work like a compiler -- figure out how to write the
rule in such a way that it is always event-driven and forward looking.
In many cases it is very natural instead to simply write a query over
the past that ``looks back'' and checks if the desired condition is
true, on demand. 

Our basic move, therefore, is to augment TCC with {\em history}.
When moving from time step $t$ to $t+1$, we propose to retain the
constraints computed at $t$, and time-stamp them with $t$.  
Thus the store will contain not just the current constraint, but also,
separately and equally, past constraints, each tagged with the time at
which they were computed. 

A simple way to accomodate this view in TCC is simply to work over
the constraint system $H(C)$ built from $C$ as follows. 
The tokens in $H(C)$ are of the form $\TIME{i}{c}$ for some $i$,
where $c$ is a token of $C$. A multiset $\Gamma$ of such tokens
entails $\TIME{k}{d}$ only if $\Gamma_k$ entails $d$ (in
$C$), where $\Gamma_k$ is the set of all constraints
$c$ such that $\TIME{k}{c} \in \Gamma$. Given $k$, by abuse of
notation we will say that $\Gamma$ entails $\PREV{c}$ at $k$ if $\Gamma_{k-1}$
entails $c$ (in $C$). Using $H(C)$, the user can write ask agents that query the past.
Tell agents must still be prevented from modifying the past by
ensuring that they can only assert constraints about the instantaneous
state. The operational semantics is now modified to carry past
constraints automatically in the constraint store, and to tag tell
constraints with the current time step. 

\begin{example}[Querying the past in TCC($H(C)$)]
The rule:
{\footnotesize
\begin{verbatim}
always if ((prev price(IBM)) > price(IBM)) then signalIBMDrop
\end{verbatim}}
\noindent will trigger if there is a drop in price of IBM stock over
successive time instants.
\end{example}
Unfortunately, this simple technique is not powerful enough. What if
we wanted to trigger a rule if the current price is less than half the
price at any point in the past when MSFT stock was above a certain
threshold? In other words, it is natural to require {\em recursive}
computations in our queries, capable of examining the past at
arbitrary depth. 

\cite{radha-fsttcs05} in fact develops such a rich framework for CCP,
called RCC. RCC is based on the idea that a judgement $A_0, \ldots,
A_{n-1} \vdash G$ can be regarded as asking whether the system of
concurrently interacting agents  $A_i$ ($i < n$) satisfy the query or
goal $G$. Appendix~\ref{sec:RCC} provides more details.)
Queries are internalized in the agent language through the production
\code{A::= if G then A}. 

Queries are not restricted to primitive constraints
\code{c}. Recursive queries are permitted. {\em Universal} queries, 
\code{G::= all V in G}, are permitted, where \code{V} is a first-order
variable. Such a query can be thought of as succeeding only when the
query \code{G} succeeds, where \code{V} is a brand-new variable that
does not occur in the agents.  {\em  Hypothetical} queries
are also permitted: \code{G::= if A then G} can be thought of as
temporarily augmenting the system currently being 
tested with $A$ and asking if the augmented system satisfies $G$. If
so, the guard is satisfied and execution continues, with the temporary
augmentation discarded.  Hypothetical queries permit ``what if''
reasoning and allow for a compact representation of very powerful
idioms. To implement this, the underlying infrastructure
must support the notion of copying the entire concurrent assembly of
agents. 

\cite{radha-fsttcs05} shows that the computational interpretation is
sound and complete with respect to the obvious logical interpretation
of the queries. 

\paragraph{TCC with deep guards.} 
We now consider how to apply these ideas to TCC. Clearly, we need to
augment the power of guards, \code{G}. To add recursion across time, we introduce
\code{G ::= hitherto G} (analogously to \code{A ::= hence A}). The
query \code{hitherto G} is intended to be true if \code{G} is true at
every point in the past (excluding the current one). We also
introduce recursive queries, \code{G ::= mu X G }, and require that \code{X} be
guarded in \code{G} (occur inside a \code{hitherto}). Similarly, we
introduce universal queries \code{G ::= all V in G}.

We could introduce hypothetical queries,\code{G::= if A
  then G}. However, we can do something richer. Note that \code{A} is
not permitted to operate in the ``past'', i.e.{} it is not possible
for an agent to spawn an agent to ``change the past''.  (Concretely,
\code{A::= hitherto A} is not allowed. This is fundamental to the
basic idea that computation always moves ahead in time.)
However, within the scope of hypothetical execution, it does make sense to
add agents to the past -- these agents are free to participate in
``what if'' reasoning, exploring what might have been. 
Therefore we introduce a new category of {\em nested agents}, \code{B},
which is the same as \code{A} except that it permits \code{B::=
  hitherto B}, and add \code{G::= if B then G}.

With these constructs it is possible for a query to ``move back in
time'' arbitrarily deeply, spawn agents in the past, and ask
queries of the modified system. Still, the nested agents and queries
are asymmetric: nested agents can move back in time
(\code{hitherto B}) as well as forward (\code{hence B}), but queries
can only move backwards.  Logically, it makes sense, then, to permit
queries to also move forward in time; we add \code{G::=
  hence G}).  This permits us to express a query that checks whether
the day after the last time IBM stock fell $10\%$ it was the case that
MSFT stock rose $10\%$. The natural formulation of this query would
involve moving back in time and then forward. 

Table~\ref{table:syntax} summarizes the language being considered,
which we name ``TCC, with history''. 
The basic picture of computation supported
by this language is as follows: The system interacts with incoming
events in a synchronous fashion. The rate at which events arrive 
is controlled by the environment and not by the system. Each
interaction marks the progression of the system down a time-line. At
each instant, the state of the system carries the entire state of past
interactions.  This is accessible to be queried in a very rich way
through a query language which permits computations to move backwards
and forwards in the past, and also spawn hypothetical queries.
However, querying cannot change the actual past, only read it. 

This paper takes the first step in studying this language. 
Section~\ref{sec:history} discusses how some
interesting  idioms can be expressed in this language. For reasons of
space we omit standard extensions of the query language with ``bag
of'' operators that permit the collection of some statistic over all
answers to a query (these are very important in practice).
Section~\ref{sec:model} formalizes the informal reasoning presented 
here. We conclude with an outline of the work that lies ahead.

\begin{table}
{\footnotesize
\begin{tabular}{l}
(Agents) A  {::=}  c   \alt \ift{G}{A}    \alt \ife{c}{A}    \alt \AND{A}{A} \alt \some{V}{A} \alt \hence{A}   \\
\quad  \alt X     \alt \MU{X}{A}   \\ \quad \\
(Goals)  G   {::=}  c   \alt \ift{B}{G}   \alt \AND{G}{G} \alt \OR{G}{G} \alt \all{V}{G} \alt \hence{G} \\
\quad \alt \hitherto{G}   \alt X     \alt \MU{X}{G}   \\ \quad \\
(Nested Agents)  B {::=}   c   \alt \ift{G}{B}    \alt \ife{c}{B}    \alt \AND{B}{B} \alt \some{V}{B} \\
 \quad \alt \hence{B} \alt \hitherto{B}  \alt X     \alt \MU{X}{B}   \\
\end{tabular}}
\vspace{5mm}

\noindent Agents $A$ are those $B$'s which do not have any occurrence
of the \code{hitherto} combinator.

\caption{TCC, with history}\label{table:syntax}
\end{table}

\paragraph{Contributions.}

The contributions of this paper may be summarized as follows:
\begin{itemize}
\item We motivate the use of TCC for complex temporal event
  processing. TCC is capable of handling the absence of information. 
\item We extend TCC with a way to capture the past history of the
  system. This permits a natural declarative style of querying the
  past. 
\item We motivate the introduction of {\em recursive queries} in
  TCC. This permits recursive queries that can reach arbitrarily
  deeply into the past.
\item Motivated by \cite{radha-fsttcs05} and the testing
  interpretation of intuitionistic logic, we further introduce
  ``hypothetical'' queries \ift{B}{G} that ask if the current system
  augmented with the agent \code{B} can answer the query \code{G}. 
  Unlike TCC, we also permit such nested agents to move backwards in
  time, allowing speculative augmentation of the past.  Together,
  these two capabilities permit \code{B} to move backwards and
  forwards in time, while confined to the past.
  \item We provide a formal operational semantics for the language,
    based on an interpretation of programs as formulas in linear time
    temporal logic, and computation as deduction.
  \item We establish that the semantics of this language is
    conservative over TCC. That is, the behaviors of a program in
    this language that is also expressible in TCC are exactly the
    same as in TCC.
\end{itemize}

\subsection{Related Work}
Several authors have explored the
properties of TCC in the last two decades, extending it in various directions. 
\cite{tini1999expressiveness} shows that the synchronous languages Lustre and Argos can be embedded in TCC. Expressiveness is further discussed in \cite{Nielsen:2002:EPT:571157.571173}: different variants that express recursiveness in different ways are discussed and related. It is shown that equivalence of programs with replication (or parameterless recursive procedures) is decidable.
\cite{Palamidessi:2001:TCC,nielsen2002temporal} propose an extension to TCC (ntcc) that can handle asynchronous communication, and nondeterministic behavior, by providing a guarded-choice operator and an unbounded but finite delay operator. A denotational semantics, and a proof system for temporal properties are presented. Another approach to reasoning about TCC programs is provided in \cite{de2001temporal,deBoer2002}. More decidability results for TCC and ntcc are presented in \cite{valencia2005decidability}: strongest post-condition equivalence for ``locally independent'' ntcc programs is shown to be decidable. This language is capable of specifying certain kinds of infinite-state reactive systems. \cite{bistarelli2008timed} discusses a variant capable of dealing with ``soft'' constraints and preferences; the intended application area is a collection of agents negotiating over quality of service.  Abstract diagnosis for a variant of TCC is considered in \cite{DBLP:journals/corr/abs-1109-1587}. 

In terms of implementation, \cite{saraswat2003jcc} describes an initial implementation in Java, for reactive computation. This is currently being extended to an implementation of the language discussed in this paper, on top of X10 \cite{vj-clock,oopsla05}.

\section{Programming in TCC with histories}\label{sec:history}
%% Discuss first just replacing C with H(C).
%% What can you express, what cant you express. Cant go back and
%% forth.
%% Hence introduce deep guards. Talk about RCC, and adapting to this setting.
We now consider how several idioms of practical interest can be
expressed in this language. 

\subsection{A concrete TCC language, V}
To fix intuitions, we work on top of a constraint system which permits
(sorted) function and predicate symbols, with equality (``\code{=}'').
Amongst the sorts available are \code{Boolean}
and \code{Int}. Sorts are closed under products and function space,
i.e. if \code{S1}, \code{S2} are sorts, then so are \code{S1 $\times$
S2} and \code{S1 => S2}. 

\vspace{5mm}
{\footnotesize
\begin{tabular}{l}
(Terms) s,t {::=} X \alt f(t1,\ldots,tn) \\
(Constraints) A,B  {::=}  s=t   \alt p(t1,\ldots, tn) \alt c,c
\end{tabular}}
\vspace{5mm}

The equality predicate is interpreted as a congruence
relation (it is symmetric, reflexive and transitive, and equal terms
can be substituted for each other in all contexts). A set of
constraints \code{c1,\ldots, cn} entails \code{p(t1,\ldots, tk)} if
and only if it entails \code{s1=t1, \ldots, sk=tk} (for some
terms \code{s1},\ldots, \code{sk}) and for some \code{i}, \code{ci}
is \code{p(s1,\ldots, sk)}. 

For convenience, we will also permit linear arithmetic constraints,
and arithmetic inequality, \code{<, <=}. 

We will also find it convenient to permit \code{prev(t)} as a term,
when \code{t} is a term. A constraint store can
establish \code{prev(u)=v} at time $t$ if it can establish
\code{u=v} at time $t-1$, and \code{v} is {\em rigid}, i.e.{} does not
change value with time. The only rigid terms are the constants -- we
assume they denote the same value at every time instant.

% Well knpown
% congruence algorithms (e.g. Shostak) can be used to implement the
% constraint system. 

We shall adopt the convention of
specifying named agents through agent clauses of the form \code{ a -:
  A}, and named goals through goal clauses of the form \code{g :-
  G}, where \code{a} and \code{g} are atomic formulas. The predicate names
for agents, goals and primitive constraints are understood as being
drawn from disjoint spaces. 

\subsection{Programming in V}

\begin{example}[\code{past G}, \code{next G}] We define the query \code{past(X=Y)}, 
  intended to be true at query time $i$ precisely if \code{X=Y} is true at
  query time $i-1$. 
{\footnotesize
\begin{verbatim}
past(X=Y) :- all U in if (hitherto hitherto U=true) then hitherto (U=true or X=Y)
\end{verbatim}}
Here is how we understand it. To establish the goal \code{past(X=Y)} in a configuration $\Gamma$ at
query time $i$, we are permitted to assume \code{U=true} (at all
times) in $[0,i-2]$, for a brand-new variable \code{U}. In turn, we
must establish either \code{U=true} or \code{X=Y} in $[0,i-1]$. 
Clearly, the assumption establishes
the desired goal in $[0,i-2]$. Hence we are
left with time $i-1$. No agent in $\Gamma$ knows about 
\code{U}. Therefore the only way \code{past(X=Y)} can be established
is if at the previous time instant \code{X=Y} can be established. 
\end{example}
The \code{past} predicate can be defined in a similar way for other
constraints of interest, e.g.:
{\footnotesize
\begin{verbatim}
past(X>Y) :- all U in if (hitherto hitherto U=true) then hitherto (U=true or X>Y)
\end{verbatim}}

The code \code{next(X=Y)} is the dual:
{\footnotesize
\begin{verbatim}
next(X=Y) :- all U in if (hence hence U=true) then hence (U=true or X=Y)
\end{verbatim}}

\begin{example}[\code{once G}] We express the query \code{once G} that succeeds
only if \code{G} can be established at some point in the past. We
illustrate for \code{G} of the form \code{X=Y}. 
{\footnotesize
\begin{verbatim}
once(X=Y) :- X=Y or past(once(X=Y)).
\end{verbatim}}
This goal can be established in a configuration at $i$ only if
\code{G} can be established at $i$, or, recursively, the goal can be
established at $i-1$. 
\end{example}

If arithmetic is available, and recursion with parameters, one can
program \code{within t do G}: 
\begin{example}[\code{within t do G}] We require \code{G} to be
  established within \code{t} time units in the past:
{\footnotesize
\begin{verbatim}
within T do X=Y :- X=Y or (T > 0 and past(within T-1 do X=Y)).
\end{verbatim}}
\end{example}

We show that the query language has enough power 
to internalize \code{else}. 
\begin{example}[\code{not(X=Y)}] We express the query \code{not(X=Y)}. This query
  succeeds only if \code{X=Y} cannot be established:
{\footnotesize
\begin{verbatim}
not(X=Y) :- all U in if (if X=Y else U=false) then U=false
\end{verbatim}}
This goal can be established in a configuration $\Gamma$ only if
\code{$\Gamma$, if X=Y else U=false} can establish \code{U=false}. But this can
happen only if $\Gamma$ can evolve in such a way that \code{X=Y} cannot
be established (per the semantics of the TCC if/else).
\end{example}

\begin{example}[\code{last X then G}] We would like to express 
  that \code{G} is true at the last time instant at which
  \code{X} was true (assuming there is a time instant at which
  \code{X} is true):
{\footnotesize
\begin{verbatim}
last X=Y then U=V :- prev last1 X=Y then U=V.
last1 X=Y then U=V :- (X=Y and U=V) or (not(X=Y) and prev(last1 X=Y then U=V).
\end{verbatim}}
Intuitively, at the last time instant, a check is made for
\code{X=Y}. If it is true, then \code{U=V} must be true, else the goal
will fail. If it is not known to be true, then the goal succeeds
provided that the same goal can be established at the previous time
instant. 
\end{example}

We turn now to using these general constructions to show how a complex
event query can be formulated.
\begin{example}
An example of the use of this goal is the query that returns the
previous price of a stock. We shall imagine that if in a time instant
an event arrives that specifies the price \code{P} of a stock \code{S}, then the 
constraint \code{price(P)=S} is added to the store. Note that many
stock price events may arrive at the same time instant -- we assume
that all are for different stocks. It is not necessary that each time
instant contains a constraint about the price of a given
stock \code{S}. In this case, we may wish to determine the previous
prices of the stock \code{S}, which is the price of the stock at the
first instant before the current one at which a price event was received.
{\footnotesize
\begin{verbatim}
prevPriceOfStock(S)=P :- 
   (prev(price(S))>0 and P=prev(price(S))) or 
   (not(prev(price(S))>0) and prev(prevPriceOfStock(S)=P)).
\end{verbatim}}
\end{example}
\noindent Now one can use this query to determine whether the price
   has dropped. The query checks that there is a price event at the
   current time instant, and the price it specifies for the
   stock \code{S} is less than the previously known price for \code{S}.
{\footnotesize
\begin{verbatim}
priceDropped(S) :- prevPriceOfStock(S) > price(S).
\end{verbatim}}
\noindent Such a query can now be used to time an agent. The agent 
{\footnotesize
\begin{verbatim}
time next^10 emitVolatilityWarning on priceDropped(S)
\end{verbatim}}
\noindent will emit a volatility warning at the tenth time instant at which the
price has dropped. Using standard TCC idioms, this agent can be
packaged up thus:

{\footnotesize
\begin{verbatim}
every hour
  do time next^10 emitVolatilityWarning
     on priceDropped(s)
  watching hour.
\end{verbatim}}

\noindent to precisely capture the rule ``Every tenth time the price drops
within an hour, emit a volatility warning''.

The above provides a flavor of the richness of this system.

\section{Semantic model}\label{sec:model}
%% Includes operational semantics.

\subsection{Transition relations}

The central problem we address is the temporal evolution of (mutually
dependent) agents and guards. 

We add a new formula $B ::= \TIME{i}B$, to keep track of formulas that
are intended to hold at a point in time in the past. We abuse notation
slightly by permitting $\TIME{0}{B}$ and treating it indistinguishably
from $B$. 
Below,
$\Gamma,\Delta,\Pi$ ranger over (possibly empty) multisets of $B$
formulas. For a multiset of formulas $\Delta=B_0,\ldots, B_{n-1}$ we
let $\TIME{i}{\Delta}$ stand for $\TIME{i}{B_0},\ldots,
\TIME{i}{B_{n-1}}$. Similary for $\hence{\Delta}$ and $\hitherto{\Delta}$.

We define three transition relations. All of them are indexed with the
{\em current} time instant $j$ and the {\em query} time instant $i$
(with $i \leq j$). The main
relation of interest is $\Gamma \proves{i}{j}{b} G$  (read: ``$\Gamma$
{\em proves} $G$ at (past) query time $i$ (with quiescent store $b$)
when the current time is $j$ ($i \leq j$)'').   We need to carry $j$
in the relation because in order to prove a goal of the form
$\hence{G}$ we only need to consider time steps upto $j$. Note that
$\Gamma$ will, in general, contain formulas active at different time
instants $k \leq j$ (i.e.{} $\Gamma$ will contain formulas of the form
$\TIME{k}{B}$). $G$ however, is never explicitly timed, since at query
time $i$ we care only about queries holding at time $i$.

To define this relation, we need two auxiliary relations that define evolution
{\em within} time instants in the past ($\Gamma\evolves{i}{j}{b}
\Gamma'$), and {\em across} time instants in the past ($\Gamma
\steps{i}{j} \Gamma'$). Note that these auxiliary relations may work
with hypothetical pasts, since they may reflect the presence
of assumptions $B$ made by the goal $G$ being solved at $j$.   

We let $\sigma^i(\Gamma)$ stand for the set of all formulas $c$ s.t.{}
 $\TIME{i}{c} \in \Gamma$, i.e.{} the subset of constraints known to
 be in effect at time $i$.

\subsubsection{The provability relation for goals}
The logical rules are straightforward, and correspond to RHS rules for
the appropriate logical connective, in a sequent-style presentation.   
Rule~\ref{eqn:constraint-G} uses $\sigma^i$ to pick out the
constraints in effect at query time $i$ from the current configuration.
Rule~\ref{eqn:ift-G} ensures that in order to prove a goal
of the form $\ift{B}{G}$ at query time $i$, the assumption $B$ is
added at time $i$ to the current configuration. 
{\footnotesize
\begin{align}
\label{eqn:constraint-G}
 \typerule
 {\sigma^i(\Gamma) \vdash c}
 {\Gamma \proves{i}{j}{b} {c}}
&  \quad \typerule
  {\Gamma \proves{i}{j}{b} {G[\mu X\,G/X]}}
  {\Gamma \proves{i}{j}{b} {\mu X\,G}}
\\
   \typerule
   {
     \begin{array}{ll}
       \Gamma \proves{i}{j}{b} G_0 & \quad \Gamma \proves{i}{j}{b} G_1 
     \end{array}
   }
   {\Gamma \proves{i}{j}{b}\AND{G_0}{G_1}}
&\quad \label{eqn:ift-G}
  \typerule{\Gamma, \TIME{i}{B} \proves{i}{j}{b} G}
  {\Gamma \proves{i}{j}{b} \ift{B}{G}} 
\\ 
  \typerule
  {\Gamma \proves{i}{j}{b} G_0}
  {\Gamma \proves{i}{j}{b} \OR{G_0}{G_1}}
& \quad
  \typerule
  {\Gamma \proves{i}{j}{b} G_1}
  {\Gamma \proves{i}{j}{b} \OR{G_0}{G_1}}
\\ 
  \typerule
  {\Gamma\proves{i}{j}{b}G[t/v]}
  {\Gamma\proves{i}{j}{b} \some{V}{G}}
& \quad
  \typerule
  {
    \begin{array}{ll}
      \Gamma \proves{i}{j}{b} {G} & \quad \mbox{($V$ not free in $\Gamma$)}
    \end{array}
  }
  {\Gamma \proves{i}{j}{b} {\all{V}{G}}}
\end{align}}

\noindent Note that $b$ is not used in these rules; we will see later
that it is used when specifying how the LHS evolves within a time instant.

 We consider now the temporal rules. These rules have a (finite) set of
assumptions, indicated by the for all quantifier. A goal
$\hitherto{G}$ can be proved at query time $i$ if it can be proven at
every time in $[0,i)$. A goal 
$\hence{G}$ can be proved at query time $i$ if it can be proven at
every time in $(i,j]$.

{\footnotesize
\begin{align}
  \typerule
  {\Gamma \proves{0}{j}{b_0} {G}}
  {\Gamma \proves{1}{j}{b} {\hitherto{G}}} &\quad
  \typerule
  {\Gamma \proves{i-2}{j}{b_0} {\hitherto{G}} \ \ldots\  \Gamma \proves{i-1}{j}{b_1} {G}}
  {\Gamma \proves{i}{j}{b} {\hitherto{G}}} \\
  \typerule
  {\Gamma \proves{i+1}{j}{b} {G} \  \Gamma \proves{i+1}{j}{b} {\hence{G}}}
%  {\forall k (i < k \leq j): \Gamma \proves{k}{j}{b} {G}}
  {\Gamma \proves{i}{j}{b} {\hence{G}}} &\quad
  \typerule
  {\Gamma \proves{j}{j}{b_0} {G}}
  {\Gamma \proves{j-1}{j}{b} {\hence{G}}}
\end{align}
}

Now, since we permit $B$'s to occur on the LHS, and these could
evolve, we must also have the following rules. In
Rule~\ref{rule:evolve-L-i}, the configuration is partitioned into
three groups of formulas -- $\Gamma_{i-}$ which are all the formulas
$\TIME{k}{B}$ with $k <i$, $\Gamma_{i+}$ which are all the formulas
$\TIME{k}{B}$ with $k\geq i$, and $\hitherto{\Delta}$.
The rule captures the notion that to prove a formula $G$ at time $i$ one must
``go back'' to time $0$ in order to account for the effects of any
$\hitherto{B}$ formulas in $\Gamma$. In this traversal into the past,
$\hitherto{B}$ agents are carried ``backwards'' (exactly as
$\hence{B}$ agents are carried forward in TCC, see
Rule~\ref{rule:evolve-hence}),
together with ``past'' state.
 The recursion is stopped by Rule~\ref{rule:evolve-L-base}. 

{\footnotesize
\begin{align} \label{rule:evolve-L-i}
\typerule
  {
    \begin{array}{llll}
     \Gamma_{i-},\TIME{i-1}{\Delta},\hitherto{\Delta} \steps{i-1}{j} \Gamma' 
     &\quad \Gamma',\Gamma_{i+} \evolves{i}{j}{b} \Gamma''
     & \quad \Gamma'' \proves{i}{j}{b} G & \quad i > 0
    \end{array}
  }
  { 
    \Gamma_{i-},\Gamma_{i+},\hitherto{\Delta} \proves{i}{j}{b} G
  }
\\
\label{rule:evolve-L-base}
\typerule
  {
    \begin{array}{ll}
     \Gamma \evolves{0}{j}{b} \Gamma'   & \quad \Gamma' \proves{0}{j}{b} G
    \end{array}
  }
  { 
    \Gamma \proves{0}{j}{b} G
  }
\end{align}
}
\subsubsection{The evolution relation}\label{sec:evolution-rules}

The rules for $\longrightarrow$ (evolution within a time instant) are as in \cite{saraswat-tdcc}, changed
in an appropriate way to consider the more general notion of execution
at possibly past time points. (The special case of TCC execution is
obtained by considering the relation $\Gamma \evolves{j}{j}{b}
\Gamma'$ and restricting ask agents to check primitive constraints.)
At query time  $i$, an agent
$\TIME{i}{\ift{G}{B}}$ can be reduced to $\TIME{i}{B}$ provided that
  the goal $G$ can be proved from the current configuration.  To
  reduce an $\TIME{i}{\ife{a}{B}}$ agent, we use the quiescent
  information $b$ (associated with query time $i$),
  as usual for Default \cc. 

{\footnotesize
\begin{align}
  \typerule
   {
       \Gamma \proves{i}{j}{b} {G} 
   }
   {
       \Gamma, \TIME{i}{\ift{G}{B}} \evolves{i}{j}{b} \Gamma,\TIME{i}{B}
     }
\\
  \typerule
   {
      b \not\vdash a
   }
   {
       \Gamma, \TIME{i}{\ife{a}{B}}\evolves{i}{j}{b} \Gamma,\TIME{i}{B}
     }
\\ 
  \typeax
  {\Gamma, \TIME{i}{\AND{B_0}{B_1}} \evolves{i}{j}{b}  \Gamma, \TIME{i}{B_0}, \TIME{i}{B_1}}
\\ 
  \typerule
  {
       \mbox{($Y$ not free in $B,\Gamma,\Pi$)}
     }
     {
       \Gamma, \TIME{i}{\some{V}{B}} \evolves{i}{j}{b} \Gamma, \TIME{i}{B[Y/V]}
     }
 \\ 
\typeax
 {
   \Gamma, \TIME{i}{\mu X\, B} \evolves{i}{j}{b}  \Gamma, \TIME{i}{B[\mu X\, B/X]}
} 
\end{align}
}
Note that there is no rule for \hitherto{B} -- it does not contribute
to instantaneous evolution, or to the step relation. It is of use
in the proves relation when moving backwards in time.

The rules for evolution across time instances are as follows. They
differ from the rules for TCC only in that the final constraint at
the previous time step is explicitly carried forward into the
configuration at the next time step ($\TIME{i}{b}$) with the
appropriate time index ($i$) to distinguish it from the constraints
that will be generated at other time steps. The first rule is used to
advance time in a query computation, the second to advance time for
the overall (top-level) computation. Below, let $\Pi$ consist of
formulas of the form $\TIME{i}{c}$ for some $i$, and
$\Gamma'$ does not contain such formulas.
{\footnotesize
\begin{align}
  \typerule
     { 
      \begin{array}{llll}
%         \TIME{i}{B},
        \Gamma \starevolves{i}{j}{b} \Gamma',\Pi,\hence{\Delta} 
         & \quad \Gamma',\Pi,\hence{\Delta} \not\evolves{i}{j}{b}   
         & \quad \sigma^i(\Pi)=b 
         & \quad i <  j
      \end{array}
    }
    {
         \Gamma \steps{i+1}{j} \Pi,\TIME{i+1}{\Delta}, \hence{\Delta} \\\\
    }
\\ \label{rule:evolve-hence}
  \typerule
     { 
      \begin{array}{llll}
         \Gamma \starevolves{i}{j}{b} \Gamma',\Pi,\hence{\Delta} 
       &  \quad \Gamma',\Pi,\hence{\Delta} \not\evolves{i}{j}{b}   
       &  \quad \sigma^i(\Pi)=b
       & \quad i=j
      \end{array}
    }
    {
         \Gamma  \steps{j+1}{j+1} \Pi,\TIME{j+1}{\Delta}, \hence{\Delta} \\
    }
\end{align}
}
%\todo{What is the relationship between $b$ and $\TIME{i}{\Gamma}$ in
%  the rule above?}
\begin{definition}[Execution]
We say that a sequence of agents
$\Gamma_0,\Gamma_1\ldots,\Gamma_n,\ldots$  is an {\em execution} if
for all $i > 0$, $\Gamma_i \steps{i+1}{i+1} \Gamma_{i+1}$.
\end{definition}

\def\restrict{\upharpoonright}
Let $\Gamma$ be a multiset of agents. Then $\Gamma\restrict i$ is the
set of all formulas $B$ such that $\TIME{i}{B}\in\Gamma$.  

\begin{proposition}[TCC+history does not change the past.]
Let $\Gamma_0,\Gamma_1\ldots,\Gamma_n,\ldots$  be an {\em execution}. 
Then for every $j > 0$ and $m,n > j$ it is the case that $\Gamma_m
\restrict j$ and $\Gamma_n \restrict j$ are multisets of constraints
that are equivalent.
\end{proposition}

The following proposition relies on the fact that a multiset of nested
TCC+history agents cannot have sub-formulas of the form $\ift{G}{B}$
(unless $G$ is a constraint), $\ift{B}{G}$, $\hence{G}$,
$\hitherto{G}$, $\hitherto{B}$.  Therefore in any proof of the
judgement $\Gamma \proves{j}{j}{b} c$ only judgements of the form
$\Gamma' \proves{j}{j}{b} c'$ are generated. No ``travel'' in time is possible.

\begin{proposition}
Suppose $\Gamma$ is a multiset of nested TCC+history agents such that
$\Gamma\restrict j$ is a multiset of TCC agents.  Then 
$\Gamma \proves{j}{j}{b} c$ iff $\Gamma\restrict j \vdash_b c$, where
$\vdash_b$ represents the TCC entailment relation with $b$ the final
resting point.
\end{proposition}

\begin{theorem}
TCC + History is conservative over TCC.
\end{theorem}

\section{Conclusion}\label{sec:conclusion}
This paper represents the first step in the study of TCC, augmented
with history and a rich notion of queries. A number of areas of work
open up.

\paragraph{Expressiveness.} 
Does this language realize the intuition that queries can be
multi-form in time, just as agents can be multi-form in time? Is it
semantically meaningful to consider deep negative guards? 

\paragraph{Denotational semantics.} The basic semantic intuition is
that this language permits rich querying of the past, with a deep
interplay between agents and guards. Since the past is not modified it
should be possible to adapt the denotational semantics of
\cite{saraswat-tdcc} (based on prefix-closed sets of traces) to this
setting. 

\paragraph{Finitary implementations.} For many uses of the language,
it would be valuable to bound the amount of past information that
needs to be carried in the state. Does this language admit of finite
state compilability (a la TCC)? If not, what restrictions need to be
placed to achieve finite state compilability?

{\footnotesize
\bibliographystyle{abbrv} %plain plainnat abbrvnat abbrv
\bibliography{x10,ddc,vldb,master,biblio,constrained-types,tcc}
}

\appendix
\section{Background}\label{sec:background}
%% Discuss TCC.
%% Retrace basic development, cc to tcc, then dcc. remark about continuous and how that is not in current scope.

The basic idea of TCC may be summarized as follows: 

\begin{quotation}
   TCC = CCP + Synchrony hypothesis 
\end{quotation}

CCP, concurrent constraint programming, is a simple view of parallel
computation that arises from multiple interacting agents sharing a
common store of constraints. Constraints are expressions (such as \code{X >=
Y + Z}) over a finite set of free variables. Each constraint is
associated with a {\em solution set}, a set of mappings from variables to
values (called {\em valuations}) that makes the constraint ``true''. e.g. the
set of valuations that makes \code{X >= Y + Z} true is the set of valuations
\code{T} s.t. \code{T(X)}, \code{T(Y)} and \code{T(Z)} are
numbers satisfying \code{T(X) >= T(Y) + T(Z)}.

Two fundamental operations on constraints are used in CCP -- \code{tell c}
(add \code{c} to the current store), and \code{ask c} (check if \code{c} is entailed by the
current store). Note that addition is conjunctive -- the solution set of
\code{c,d} is the intersection of the solution sets of \code{c} and \code{d}. Say
that \code{c} {\em entails} \code{d} if the solution set of \code{c} is contained in that of \code{d}
(that is, if \code{v} is a solution for \code{c}, then it is a solution for \code{d}) and
{\em disentails} \code{d} if the solution sets of \code{c} and \code{d} are disjoint.  
The operation \code{ask c} succeeds if the store entails \code{c}, fails if
the store disentails \code{c}, and suspends otherwise.  

In CCP, the programmer specifies a set of agents over shared variables
that interact with each other by telling and asking constraints on the
shared variables. The fundamental property of CCP is that computations
are determinate -- the result is the same, regardless of the order in
which agents are executed. Furthermore, programs have a declarative
interpretation, they can be read as formulas in logic and have the
property that if a program \code{P} logically entails a constraint \code{c}, then
execution of \code{P} will result in a store that entails \code{c}. 

CCP is a rich and powerful framework for (asynchronous) concurrent
computation.  TCC arises from CCP by ``extending'' CCP across time. We
add the new control construct \code{next}: if \code{A} is an agent, then so is
\code{next A}. The intuitive idea is that computation progresses in a series
of steps. In each step, some input is received from the environment
(an ``event"), and added to the store. The program is then run to
quiescence. This will yield a store of constraints, this provides the
``instantaneous response". In addition it will yield a set of next \code{A1},
\ldots, \code{next An} agents. (Note some of these agents can be simple
constraints.) These are precisely the agents that are used to respond
to the next event, at the next time instant.

Notice that this view is concerned with a logical notion of time --
time is just a sequence of ticks arriving from the environment (with
additional input). There is no intrinsic association of this sequence
of ticks with ``real" time, e.g. msecs. This is the powerful insight
that underlies the notion of multiform time. This notion says that the
temporal constructs in the language can all be used for any
user-defined notion of time, not just the ``built-in" notion of time.
In TCC, this is captured by the \code{time A on B} combinator. For the agent
\code{A}, the agent \code{B} defines the notion of time -- only those time ticks
that ``pass" the test \code{B} are passed on to \code{A}. Thus \code{A} is executed with a
``programmer supplied" clock. Of course, these constructs can be
nested, thus time \code{time time A on B1 on B2} will supply to A only those time
ticks that pass \code{B1} and \code{B2}.

This flexibility of the basic formalism permits a large number of
combinators to be definable by the user. Combinators such  as the
following are definable in \code{A} and \code{B}: 
\begin{description}
\item[\code{do A watching c}:]
   Execute \code{A}, across time instants  but abort it as soon as there is a
   time instant which satisfies the constraint \code{c}

\item[\code{supend c activate d A}:]
   Execute \code{A}, across time instants, suspending it as soon as a time
   instant is reached in which \code{c} is true. Then activate it as soon as
   a time instant is reached in which \code{d} is true.
\end{description}

\subsection{RCC-- Combining agent execution and testing}
\newcommand\Entails[2]{\withmmode{#1 \Rightarrow #2}}
\newcommand\Some[2]{\texttt{some}\;#1\; \texttt{in}\; #2}
\newcommand\All[2]{\texttt{all}\;#1\; \texttt{in}\; #2}
\newcommand\CCPAnd[2]{\texttt{all}\;#1\; \texttt{and}\; #2}
\newcommand\Or[2]{\texttt{all}\;#1\; \texttt{or}\; #2}
The key intuition was the recognition that CCP corresponds to
``computation  on the left'', or forward chaining, and (definite
clause) logic programming corresponds to backward chaining.  This is
illustrated by the following characterization of CCP agents as
formulas in intuitionistic logic:

{\footnotesize
\begin{tabular}{llll}
(Agents) & A &{::=} & c \alt \Entails{G}{A} \alt E \alt \Some{V}{A} \\
(Goals) & G &{::=} & c \alt \CCPAnd{G}{G} \\
(Clauses) & P & {::=} & \Entails{E}{D} \alt \CCPAnd{P}{P}
\end{tabular}}

\noindent Computation is initiated on the presentation of an initial agent,
\texttt{A}, and progresses in the ``forward'' direction. One thinks of
a sequent $\texttt{A}_1, \texttt{A}_n \rightarrow$ as a multiset of interacting
agents operating on a store of constraints (the subset of the
$\texttt{A}_i$ that are constraints). If the store is powerful enough
to entail the condition \texttt{G} of an agent \Entails{G}{A}, then 
\Entails{G}{A} can be replaced by \texttt{A}. This corresponds to the
application of the left hand rule for implication. Recursive calls
\texttt{E} are replaced by the body \texttt{A} of their defining
clauses \Entails{E}{D}. Computation terminates when no more
implication can be discharged. 

This is logically sound. Clearly if we start computing with an agent
\texttt{A} and terminate in a state with the subset of constraints
$\sigma$ then we have $A\vdash \sigma$, where $\vdash$ represents
provability in Intuitionistic Logic (IL), augmented with axioms from
the underlying constraint system, $\cal C$.  Is this logically
complete? Indeed -- \cite{saraswat92higherorder} shows that if there is a
constraint \texttt{d} that is entailed by \texttt{A}, then in fact it
is entailed by $\sigma$ the constraint store of the final
configuration obtained by executing \texttt{A} as a CCP agent. Hence
CCP operational semantics is sound and complete with respect to
entailment of constraints. 

Note that this language corresponds to ``flat'' guards.
In the early development of concurrent logic programming languages
\cite{Shapiro-CP,saraswat-popl87,GHC} a lot of attention was paid to ``deep''
guards. How can deep guards be integrated into CCP?

One idea is to look at definite clause logic programming. The logical
picture here is well known. 

{\footnotesize
\begin{tabular}{llll}
(Goals) & G &{::=} c \alt \CCPAnd{G}{G} \alt \Or{G}{G} \alt H \alt \Some{V}{G}\\
(Clauses) & P & {::=} \Entails{H}{G} \alt \CCPAnd{P}{P}
\end{tabular}}

\noindent Computation corresponds to posing a query, or a goal, against a
database of clause of the form \Entails{H}{G} \cite{CLP}.  A configuration
consists of a collection of \texttt{G} formulas.  In each step, a conjunction
is replaced by its components, an existential \Some{V}{G} by
\texttt{G}, with \texttt{V} a ``new''  variable, and an atom
\texttt{H} by the body \texttt{G} of a clause \Entails{H}{G} from the
program. A disjunct is non-deterministically replaced by one of its
disjuncts. Computation terminates when the configuration contains only
constraints.  Of particular interest are terminal configurations in
which the constraints are jointly satisfiable, these correspond to
answers for the original query.

Is there a reasonable way to combine the two?  In fact, it is possible
to do this, and a lot more.  It is possible to give an intuitive
operational semantics for the following system of agents and goals. 

{\footnotesize
\begin{tabular}{llll}
(Agents) & D &{::=} & c 
   \alt \Entails{G}{D} 
   \alt E 
   \alt \Entails{E}{D} 
   \alt \CCPAnd{D}{D} 
   \alt \Some{V}{D} 
    \alt \All{V}{D}\\
(Goals) & G &{::=} & c 
    \alt \Entails{A}{G} 
    \alt H 
    \alt \Entails{G}{H}
    \alt \CCPAnd{G}{G} 
    \alt \Or{G}{G} 
     \alt \Some{V}{G} \\
&&&     \alt \All{V}{G}\\
(Clauses) & P & {::=} & \Entails{H}{G} \alt \Entails{E}{D} \alt \CCPAnd{P}{P}
\end{tabular}}

\noindent Note that richness of interplay between agents and goals --
agents can be defined in terms of goals, and goals can be defined in
terms of agents. 

What is the underlying programming intuition? We think of 
\texttt{D} as representing a concurrent, interacting system of agents
(interacting through a shared constraint store). We think of
\texttt{G} as a {\em test} of such a system. We think of a sequent $\texttt{D}
\vdash \texttt{G}$ as establishing that the system \texttt{D} {\em passes} the test
\texttt{G}.  With this interpretation, we can think of an agent
\Entails{G}{D} as saying: if the current system of agents can pass the
test \texttt{G}, then reduce to \texttt{D}. Conversely,  one thinks of
the goal \Entails{D}{G} as a ``what if'' test: Suppose the existing system
is augmented with the agent \texttt{D}. Does it now pass the test
\texttt{G}?  Similarly, \All{V}{G} is a {\em generic} goal: it asks
the question ``Does the system pass the test \texttt{G}'' for some
completely unknown variabe \texttt{V} (hence for all possible values
of \texttt{V}). 

We showed further that this semantics is sound and complete with
respect to the interpretation of agents and goals as formulas in
Intuitionistic Logic (IL). The key insight is to ``segregate'' the
atomic formulas that occur in agents (\texttt{E}) and in goals
(\texttt{H}) -- these must come from disjoint vocabularies. Therefore
the ``left hand side'' (LHS) and the ``right hand side'' (RHS) of a
sequent can no longer communicate through the application of identity
rules ($\Gamma, A \vdash A$). Rather the replacement constraint
inference rule must be used. Computation can be performed in
potentially arbitrary combinations of LHS steps and RHS steps,
corresponding to evolution of the concurrent agent and simplifcation
of the test, respectively. 

Subsequent work by Liang and Miller \cite{Liang07focusingand} established a connection between
\cite{radha-fsttcs05} and the notion of {\em focussing} proofs developed by
Andreoli \cite{Andreoli92logicprogramming}. Indeed, the notion of combining forward and
backward chaining in the very flexible way described above has seen
significant recent work.

\end{document}